\documentclass[conference,10pt]{IEEEtran}
    \usepackage{booktabs}
    \usepackage{diagbox}
    \usepackage{makecell}
    \usepackage{array}
    \usepackage{graphicx,amssymb,amsmath}
    \usepackage{multicol}
    \usepackage[noadjust]{cite} 
    \usepackage{setspace}
    \usepackage{graphicx}
    \usepackage{subfigure}
    \usepackage{float}
    \usepackage {url}
    \usepackage{stfloats}
    \usepackage{amsthm,pifont}
    \usepackage{flushend}
    \usepackage{cases,subeqnarray}
    \usepackage{bm,multirow,bigstrut}
    \usepackage{amsmath, amsthm, amssymb}
    \usepackage{textcomp}
    \usepackage{latexsym,bm}
    \usepackage{booktabs}
    \usepackage{mathtools}
    \usepackage{dsfont}
    \usepackage{extarrows}
    \usepackage{epsfig}
    \usepackage{epsfig}
    \usepackage{epstopdf}
    \usepackage[table]{xcolor}
    \usepackage{colortbl} 
    \usepackage{multirow, hhline}
    \usepackage{xcolor}
    \usepackage{array} 
    \usepackage{algorithmicx,algorithm}
    \usepackage{hyperref}
    \usepackage{caption}
    \usepackage{graphicx}
    \usepackage{float} 
    \usepackage{subcaption}

    \theoremstyle{plain}

    \theoremstyle{plain}

    \usepackage{amsmath}

    \IEEEoverridecommandlockouts
    \begin{document}
    \title{Joint Model Caching and Resource Allocation in Generative AI-Enabled Wireless Edge Networks}
    
    \author{
    Zhang Liu\textsuperscript{1},
    Hongyang Du\textsuperscript{2},
    Lianfen Huang\textsuperscript{1},
    Zhibin Gao\textsuperscript{3},
    and Dusit Niyato\textsuperscript{4},~\IEEEmembership{Fellow,~IEEE}
    \\[1ex]
    \textsuperscript{1}Xiamen University, \textsuperscript{2}The University of Hong Kong, \\ \textsuperscript{3}Jimei University, \textsuperscript{4}Nanyang Technological University
    \\[1ex]

    \text{Email: zhangliu@stu.xmu.edu.cn, duhy@eee.hku.hk, lfhuang@xmu.edu.cn,}\\ 
    \text{gaozhibin@jmu.edu.cn, dniyato@ntu.edu.sg}
    \thanks{The work was supported in part by the National Natural Science Foundation of China under Grant 62371406, Grant 61971365, and Grant 62171392, in part by the Natural Science Foundation of Fujian Province of China under Grant 2021J01004, in part by the China Scholarship Council program under Grant 202206310094, in part by the National Research Foundation, Singapore, and in part by Infocomm Media Development Authority under its Future Communications Research \& Development Programme, Defence Science Organisation (DSO) National Laboratories under the AI Singapore Programme under Grant FCP-NTU-RG-2022-010 and Grant FCP-ASTAR-TG-2022-003, in part by Singapore Ministry of Education (MOE) Tier 1 under Grant RG87/22, and in part
    by the NTU Centre for Computational Technologies in Finance (NTU-CCTF). (Corresponding author:
    Lianfen Huang.)
}
    }
\maketitle
    \begin{abstract}
    With the rapid advancement of artificial intelligence (AI), generative AI (GenAI) has emerged as a transformative tool, enabling customized and personalized AI-generated content (AIGC) services. However, GenAI models with billions of parameters require substantial memory capacity and computational power for deployment and execution, presenting significant challenges to resource-limited edge networks. In this paper, we address the joint model caching and resource allocation problem in GenAI-enabled wireless edge networks. Our objective is to balance the trade-off between delivering high-quality AIGC and minimizing the delay in AIGC service provisioning. To tackle this problem, we employ a deep deterministic policy gradient (DDPG)-based reinforcement learning approach, capable of efficiently determining optimal model caching and resource allocation decisions for AIGC services in response to user mobility and time-varying channel conditions. Numerical results demonstrate that DDPG achieves a higher model hit ratio and provides superior-quality, lower-latency AIGC services compared to other benchmark solutions.
    \end{abstract}
    \IEEEpeerreviewmaketitle
    \section{Introduction}
    \subsection{Background and Overview}
    Recent advancements in generative artificial intelligence (GenAI) have revolutionized the capabilities of traditional AI, which primarily focuses on discriminative tasks, pattern recognition, and decision-making~\cite{cao2023comprehensive}. Technically, GenAI excels at creating a diverse array of AI-generated content (AIGC), such as images, music, and natural language, based on human-provided instructions by learning the underlying patterns and distributions of training data. For instance, ChatGPT\footnote{\href{https://openai.com/index/hello-gpt-4o/}{https://openai.com/index/hello-gpt-4o/}} is a large language model that can effectively understand and generate human-like text, responding interactively to given prompts. Additionally, DALL$\cdot$E-3\footnote{\href{https://openai.com/index/dall-e-3/}{https://openai.com/index/dall-e-3/}} is an image generation model capable of creating high-quality images from textual descriptions. The remarkable achievements in GenAI have greatly enhanced productivity across various domains, such as art, advertising, and education.

    
    Nowadays, benefiting from the growth of data and model size, GenAI can capture more comprehensive distributions that closely align with reality, resulting in more realistic and high-quality AIGC. However, these large foundation model architectures necessitate substantial storage capacity for caching and extensive computational resources for inference, placing significant pressure on practical deployment. For example, GPT-3 has 175 billion parameters and requires 8 × 48GB A6000 GPUs to perform inference for generating responses~\cite{xiao2023smoothquant}. These challenges are further exacerbated in the emerging Metaverse paradigm, which demands continuous AIGC from resource-limited personal devices. Consequently, the growing footprint and complexity of GenAI models, along with extensive resources required for deployment, extremely limit the accessibility of AIGC service provision.

    Fortunately, cloud data centers with powerful computing resources, such as Amazon Lightsail\footnote{\href{https://aws.amazon.com/lightsail/}{https://aws.amazon.com/lightsail/}} and Alibaba Cloud\footnote{\href{https://www.alibabacloud.com/}{https://www.alibabacloud.com/}}, are capable of offering cloud-assisted AIGC services to users through the core network. However, migrating AIGC services to the cloud not only leads to prohibitive traffic congestion on backhaul links~\cite{10736570} but also raises concerns about privacy exposure over public cloud infrastructure. As an alternative, deploying GenAI models at the wireless edge (i.e., on edge servers co-located with base stations or roadside units) has proven to be a promising solution. First, due to the physical proximity between edge servers and users, edge-assisted AIGC service provisioning significantly reduces end-to-end latency. Second, keeping sensitive data local rather than transmitting it to cloud data centers enhances privacy protection, addressing growing concerns around data security.

    \subsection{Motivation and Main Challenges}
     Although provisioning AIGC services at wireless edge networks offers several advantages, there remain significant issues to address. \emph{First, unlike traditional content caching, caching GenAI models requires the careful coordination of both storage and computing resources.} Researchers in~\cite{9239909} investigated a content caching and user association problem in heterogeneous cellular networks, proposing a cubic exponential smoothing algorithm to determine the content cached in base stations (BSs). Researchers in~\cite{9387141} proposed a freshness-aware content refreshing scheme for mobile edge caching systems, where the BS refreshes cached content items based on the age of information (AoI) upon user requests. However, compared to content caching, GenAI model caching requires substantial computing capabilities for inference, complicating resource allocation schemes as edge servers must balance both storage and computational resources -- where larger models offer better performance but demand more resources to run.

    \emph{Second, conventional optimization methods are limited by both high computational complexity and the need for complete information.} Researchers in~\cite{tran2018joint} proposed a holistic strategy for joint task offloading and resource allocation in mobile edge computing networks, where the original problem was decomposed into an RA sub-problem and a TO sub-problem, each solved using quasi-convex and convex optimization techniques, respectively. Researchers in~\cite{gao2021resource} investigated a resource allocation problem for latency-aware federated learning in industrial Internet of Things systems, proposing a heuristic algorithm to select appropriate devices that balance the trade-off between interference and convergence time. However, convex optimization methods require extensive computations, making it difficult to meet real-time decision-making requirements. Additionally, conventional methods rely on precise knowledge of the system state to formulate and solve the optimization problem, which is often unavailable in practical networks.

\subsection{Summary of Contributions}
Motivated by these challenges, this paper investigates the problem of joint GenAI model caching and resource allocation in wireless edge networks, with the goal of delivering low-latency, high-quality AIGC services to diverse users. The contributions of this paper are summarized as follows:

\begin{itemize}
\item We formulate the joint model caching and resource allocation problem in GenAI-enabled wireless edge networks as a mixed-integer nonlinear programming (MINLP) problem, which is known to be NP-hard and particularly challenging to solve.

\item To address this problem, we employ a deep deterministic policy gradient (DDPG)-based reinforcement learning approach. After training, DDPG can efficiently map the problem’s state space (e.g., user mobility and varying AIGC requests) to optimal model caching and resource allocation decisions for AIGC service provisioning.

\item We assess the effectiveness of our approach through experiments across diverse simulation settings, showing that DDPG achieves a superior model hit ratio and provides higher-quality, lower-latency AIGC services in comparison to benchmark methods.
\end{itemize}

   \section{System Model}
    \subsection{Network Outline}
\begin{figure}[t!]
\includegraphics[width=.48\textwidth]{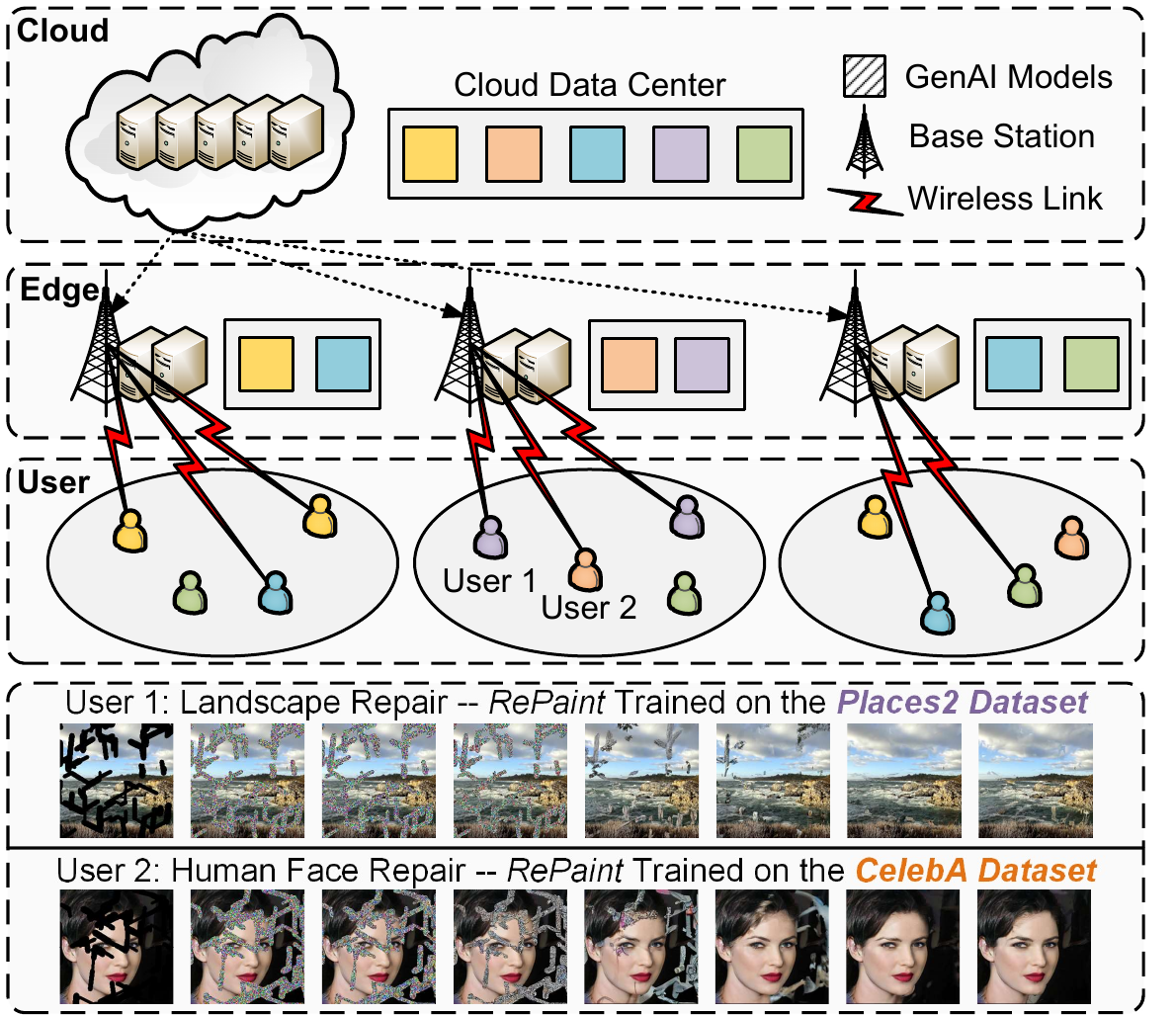}
\centering
\vspace{-1.5mm}
\caption{A schematic of the user-edge-cloud orchestrated architecture for provisioning image-generating AIGC services.}
\label{fig:Networks_Architecture}
\vspace{-3mm}
\end{figure}

    We consider a user-edge-cloud orchestration architecture, as shown in Fig.~\ref{fig:Networks_Architecture}, comprising a cloud data center with abundant computing resources for training various GenAI models, a base station (BS) with an embedded edge server responsible for caching these models, and $N$ users, represented by the set $\mathcal{N}=\{1,\cdots,N\}$. Additionally, we denote $\mathcal{M}=\{1,\cdots,M\}$ as the set of different types of GenAI models. In this paper, we focus on the diffusion-based GenAI model, \emph{RePaint}~\cite{lugmayr2022repaint}, trained on $M$ diverse datasets to represent distinct GenAI models. For example, \emph{RePaint} trained on a landscape scene dataset is suited for generating scenery images (i.e., user 1), while \emph{RePaint} trained on a dataset of celebrity faces should be used to repair corrupted human images (i.e., user 2).

    Taking into account user mobility and time-varying channel conditions, we use a discrete time representation of the system, with the time index denoted by $t \in \mathcal{T}=\{1,\cdots,T\}$. At each time slot $t$, each user generates an image-generating AIGC service request, represented by a two-tuple $[\varphi_{n}(t),d^{in}_{n}(t)]$, where $\varphi_{n}(t)$ denotes the type of AIGC service and $d^{in}_{n}(t)$ represents the input data size (in bits) of user $n$ at time slot $t$. We assume that users lack sufficient storage to cache GenAI models and must either offload the AIGC service request to the BS (if the corresponding GenAI model is cached) or to the cloud data center (which stores all GenAI models).

\subsection{Model Caching}
\vspace{-.15mm}
Given that the BS can only respond to AIGC service requests when the corresponding GenAI models are trained on the relevant dataset, and that the resource-limited edge server can only cache a subset of these models, careful orchestration of GenAI model caching decisions is essential. Specifically, at the beginning of each time slot $t$, the BS updates its model caching decisions $\bm{\varrho}(t)=\{\varrho_1(t),\dots,\varrho_m(t),\dots,\varrho_M(t)\}$, where $\varrho_m(t)=1$ indicates that the $m$-th GenAI model is cached at the BS during time slot $t$; otherwise, $\varrho_m(t)=0$. Meanwhile, we assume that the probability of user $n\in \mathcal{N}$ requesting AIGC service $m\in \mathcal{M}$ at time slot $t$ follows a Zipf distribution~\cite{yao2023cooperative}, given by
\vspace{-1.5mm}
\begin{equation}\label{eq:Zipf_distribution}
\text{Pr}\{\varphi_{n}(t)=m\} = \frac{m^{-\gamma(t)}}{\sum_{i\in\mathcal{M}}i^{-\gamma(t)}},
\vspace{-.5mm}
\end{equation}
where $\gamma(t)$ denotes the skewness of popularity at time slot $t$. 

\subsection{Communication Model}
\vspace{-.15mm}
Since users need to both send their AIGC requests to the BS and retrieve the corresponding images, we formulate the communication model from the perspectives of service migration (i.e., uplink transmission) and result feedback (i.e., downlink transmission).

\subsubsection{Service Migration} 
The transmission rate from user $n$ to the BS at time slot $t$, denoted by $R^{up}_{n}(t)$, is given by
\vspace{-1mm}
\begin{equation}\label{eq:uplink_transrate}
R^{up}_{n}(t)= b_{n}(t)W^{up}\text{log}_2\Big(1+\frac{p_nh_{n}(t)}{U_0b_{n}(t)W^{up}} \Big).
\vspace{-.5mm}
\end{equation}
Here, $W^{up}$ (in Hz) denotes the uplink channel bandwidth, $b_{n}(t)$ represents the bandwidth allocation ratio for user $n$ at time slot $t$, $p_n$ (in dBm) is the transmit power of user $n$, and $U_0$ is the noise power spectral density (in dBm/Hz). Additionally, $h_{n}(t)= g_{n}(t)|\delta_{n}(t)|^2$~\cite{lin2021semi} represents the channel gain between user $n$ and the BS at time slot $t$, capturing both large-scale path loss and small-scale signal fading. Specifically, $\delta_{n}(t) \sim \mathcal{CN}(0,1)$ is the Rayleigh fading component, varying independently across time slots, and the path loss $g_{n}(t)$ (in dB) is modeled as 
\vspace{-1.5mm}
\begin{equation}\label{eq:path_loss}
g_{n}(t)=-128.1-37.6\text{log}_{10}\text{dis}_{n}(t),
\vspace{-.5mm}
\end{equation}
where $\text{dis}_{n}(t)$ denotes the Euclidean distance between user $n$ and the BS at time slot $t$. 

Consequently, the uplink transmission delay for user $n$ to transmit its AIGC service request to the BS at time slot $t$, denoted by $D^{up}_{n}(t)$, can be calculated as follows:
\vspace{-1mm}
\begin{equation}\label{eq:uplink_transmission_delay}
\hspace{-2mm}D^{up}_{n}(t)\!=\! \left\{ \begin{array}{l}
            {d^{in}_{n}(t)}/{R^{up}_{n}(t)}, \quad \text{if}\ \varrho_{m}(t)=1, m=\varphi_{n}(t) \\
			{d^{in}_{n}(t)}/{R^{up}_{n}(t)}+{d^{in}_{n}(t)}/{R^{bc}}, \quad \text{otherwise},
		\end{array} \right.
\vspace{-.5mm}
\end{equation}
where $\varrho_{m}(t) = 1, m = \varphi_{n}(t)$ indicates that the GenAI model corresponding to the AIGC service requested by user $n$ at time slot $t$ is cached at the BS. If not, the AIGC service will be offloaded to the cloud data center, resulting in additional backhaul transmission delay, with $R^{bc}$ denoting the fixed wired transmission rate between the BS and the cloud.

\subsubsection{Result Feedback} Similar to \eqref{eq:uplink_transmission_delay}, we model the transmission rate from the BS to user $n$ at time slot $t$ as follows:
\vspace{-1mm}
\begin{equation}\label{eq:uplink_transrate}
R^{dw}_{n}(t)= W^{dw}\text{log}_2\Big(1+\frac{p_{b}h_{n}(t)}{N_0W^{dw}} \Big),
\vspace{-.75mm}
\end{equation}
where $p_{b}$ represents the BS transmit power (in dBm), and $W^{dw}$ (in Hz) denotes the downlink channel bandwidth. Consequently, the result feedback delay for user $n$ to retrieve the generated image at time slot $t$ is given by
\begin{equation}\label{eq:downlink_transmission_delay}
D^{dw}_{n}(t)= \left\{ \begin{array}{l}
            {d^{op}_{m}}/{R^{dw}_{n}(t)}, \quad \text{if}\ \varrho_{m}(t)=1, m=\varphi_{n}(t) \\[.3em]
			{d^{op}_{m}}/{R^{dw}_{n}(t)}+{d^{op}_{m}}/{R^{cb}}, \quad \text{otherwise},
		\end{array} \right.
\vspace{-.5mm}
\hspace{-3mm}
\end{equation}
where $d^{op}_{m}$ is the size of the reconstructed image (in bits) of the $m$-th type AIGC service, determined by the inherent architecture of the $m$-th GenAI model, and $R^{cb}$ is the fixed wired transmission rate between the cloud and the BS.

\subsection{Computing Model}\label{subsec:comp_model}
\vspace{-.15mm}
Compared to existing works~\cite{9239909, 9387141, tran2018joint, gao2021resource}, which primarily optimize task execution delay and energy consumption using explicit mathematical models, evaluating the performance of image-generating AIGC services presents three major challenges:

\begin{itemize}
\item \emph{What is the appropriate evaluation metric for image-generating AIGC services?}

\item \emph{How should we define the computational resources for diffusion-based GenAI models?}

\item \emph{What are the mathematical relationships between the computational resources of diffusion-based GenAI models, image generation delay, and the quality of the generated image?}
\end{itemize}

To address these questions, we first use total variation (TV)~\cite{ding2020image} to assess the perceived quality of AI-generated images, as TV quantifies ‘roughness’ or ‘discontinuity’ by summing the absolute differences between adjacent pixels; a lower TV value represents higher image quality. We then define denoising steps as the computational resource for diffusion-based GenAI models, as additional denoising steps allow for more gradual refinement and information reconstruction, improving the quality of AI-generated images. Finally, in our previous work~\cite{liu2024two}, we conducted experiments with \emph{RePaint} trained on the CelebA-HQ dataset (containing images of celebrities’ faces) and used fitting methods to develop one of the first mathematical functions that connect image generation quality, image generation delay, and computational resource consumption (i.e., denoising steps). The corresponding computing models are formulated below.

\subsubsection{Image Generation Quality} Based on the TV value, the mathematical function representing the perceived quality of the generated image for user $n$ at time slot $t$, in relation to the allocated computational resources (i.e., denoising steps), is given by~\cite{liu2024two}
\begin{align}\label{eq:image_generation_quality}
 &B^{gt}_{n}(t)\!=\! \left\{ \begin{array}{l}
            \!\!\!A_2,  \quad\quad x_{n}(t)L\leq A_1  \\
            \!\!\!\frac{A_4-A_2}{A_3-A_1}(x_{n}(t)L-A_1)+A_2, \ A_1< x_{n}(t)L< A_3 \\
			\!\!\!A_4, \quad\quad x_{n}(t)L \geq A_3
		\end{array} \right. \nonumber \\ 
  &\quad\quad\quad\quad\quad\quad\quad\quad\quad\quad\quad\quad\quad \varrho_{m}(t)=1, m=\varphi_{n}(t).
\end{align}
Here, $L$ denotes the total number of denoising steps performed at the BS, and $x_{n}(t) \in [0,1]$ represents the denoising step allocation ratio for user $n$ at time slot $t$. Additionally,~\eqref{eq:image_generation_quality} involves four key parameters: $A_1 = 60$, indicating the minimum denoising steps where image quality begins to improve; $A_2 = 110$, representing the original image quality; $A_3 = 170$, showing the threshold where image quality stabilizes, no longer improving with additional denoising steps; and $A_4 = 28$, representing the highest image quality value. Conversely, if the GenAI model corresponding to the AIGC service requested by user $n$ at time slot $t$ is not cached at the BS, we set $B^{gt}_{n}(t) = A_4$, as the cloud has sufficient computational resources to produce an image of the highest quality.

\subsubsection{Image Generation Delay} Meanwhile, the mathematical function relating the image generation time for user $n$ at time slot $t$ to the allocated denoising steps is given as follows~\cite{liu2024two}:
\vspace{-1.5mm}
\begin{align}\label{eq:image_generation_delay}
    D^{gt}_{n}(t) = B_1x_{n}(t)L+B_2, \ \varrho_{m}(t)=1, m=\varphi_{n}(t),
    \vspace{-.5mm}
\vspace{-.5mm}
\end{align}
where $B_1 = 0.18$ and $B_2 = 5.74$. Conversely, if the GenAI model corresponding to the AIGC service requested by user $n$ at time slot $t$ is not cached at the BS, we set $D^{gt}_{n}(t) = B_1A_3+B_2$, as the cloud allocates the minimum resources required to generate the image at the highest quality. Summing~\eqref{eq:uplink_transmission_delay},~\eqref{eq:downlink_transmission_delay}, and~\eqref{eq:image_generation_delay}, the total AIGC service provisioning delay for user $n$ at time slot $t$ is given by
\vspace{-0.5mm}
\begin{align}\label{eq:total_delay}
    D^{tl}_{n}(t) = D^{up}_{n}(t)+D^{dw}_{n}(t)+D^{gt}_{n}(t).
\vspace{-5mm}
\end{align}

Consequently, to achieve low-latency, high-quality AIGC services, we define a utility function as the weighted sum of generation delay and the TV value of the generated image for user $n$ at time slot $t$:
\vspace{-0.75mm}
\begin{align}\label{eq:weighted_sum}
    G_{n}(t) = \alpha D^{tl}_{n}(t)+(1-\alpha)B^{gt}_{n}(t),
    \vspace{-.5mm}
\vspace{-2mm}
\end{align}
where, $\alpha \in [0,1]$ represents a preference weight factor.

\vspace{-1mm}
\section{Problem Formulation} \label{sec:probblem_formulation}
\vspace{-.7mm}
We now formulate the joint model caching and resource allocation problem in GenAI-enabled wireless edge networks as a dynamic, long-term optimization. Our objective is to minimize the utility in~\eqref{eq:weighted_sum} across all users and time slots. This requires effective scheduling of GenAI model caching, along with computing and communication resource allocation. Mathematically, we define this problem as $\mathbb{P}$ below:
\vspace{-1.5mm}
\begin{align}
     \hspace{-3mm}&\mathbb{P}: \quad \min_{\bm{\varrho},\bm{b}, \bm{x}} \frac{1}{TN}\sum_{t\in\mathcal{T}}\sum_{n\in\mathcal{N}} G_{n}(t)\hspace{-40mm}\label{eq:problem} \\[-.45em]
     &\hspace{-3mm} \textrm{s.t.}\nonumber\\ 
     &\hspace{-3mm}  \varrho_m(t) \in \{0,1\}, \ \forall t\in \mathcal{T}, m\in \mathcal{M}, \tag{11a} \label{eq:Caching_Constraint}\\[-.2em]
     &\hspace{-3mm}  b_{n}(t) \in [0,1], \ \forall t\in \mathcal{T}, n\in\mathcal{N}, \tag{11b} \label{eq:Comm_Constraint}\\[-.2em]
     &\hspace{-3mm}  x_{n}(t) \in [0,1], \ \forall t\in \mathcal{T}, n\in\mathcal{N}, \tag{11c} \label{eq:Comp_Constraint}\\[-.2em] 
     &\hspace{-3mm}  \sum_{m \in \mathcal{M}}\varrho_m(t)c_m \leq C,  \ \forall t\in \mathcal{T}, \tag{11d} \label{eq:Caching_Resource_Constraint}\\[-.2em]
     &\hspace{-3mm}   \sum_{n \in \mathcal{N}}b_{n}(t)\leq 1,  \ \forall t\in \mathcal{T}, \tag{11e}\label{eq:Comm_Resource_Constraint}\\[-.2em]
     &\hspace{-3mm}  \sum_{n \in \mathcal{N}}x_{n}(t)\leq 1,  \ \forall t\in \mathcal{T}, \tag{11f} \label{eq:Comp_Resource_Constraint}
     \vspace{-1.5mm}
\end{align}
where $\bm{\varrho}=\{\varrho_m(t)\}_{m\in\mathcal{M},t\in \mathcal{T}}$ is the caching decision vector for GenAI models, $\bm{b}=\{b_{n}(t)\}_{n \in \mathcal{N},t\in \mathcal{T}}$ is the communication resource allocation ratio vector for users, and $\bm{x}=\{x_{n}(t)\}_{n \in \mathcal{N},t\in \mathcal{T}}$ is the computing resource allocation ratio vector for users.

In $\mathbb{P}$, constraint~\eqref{eq:Caching_Constraint} enforces a binary requirement for the model caching decision. Constraints~\eqref{eq:Comm_Constraint} and \eqref{eq:Comp_Constraint} set the valid ranges for the communication and computation resources allocated by the BS to various users. Constraints~\eqref{eq:Caching_Resource_Constraint}-\eqref{eq:Comp_Resource_Constraint} outline the limitations on the BS’s caching, bandwidth, and computing capacities. In constraint~\eqref{eq:Caching_Resource_Constraint}, $c_m$ specifies the storage needed (in GB) for the $m$-th type GenAI model, while $C$ indicates the BS’s maximum storage capacity (in GB). Clearly, problem $\mathbb{P}$ is a mixed-integer nonlinear program (MINLP) and is generally classified as NP-hard. Consequently, finding an efficient solution to problem $\mathbb{P}$ poses a significant challenge.


\section{Deep Deterministic Policy Gradient-Based Reinforcement Learning Approach}
    



\subsection{MDP Elements in the DDPG Framework}
We formulate problem $\mathbb{P}$ as a Markov Decision Process (MDP), which comprises the \emph{state space}, \emph{action space}, and \emph{reward function}.

\subsubsection{State Space} At the beginning of each time slot $t$, the DRL agent (e.g., BS) observes the state $\bm{s}(t)$ to acquire environment information, which consists of the following elements:
\begin{align}\label{eq:DDPG_state}
    \bm{s}(t)=\{\gamma(t),\bm{h}(t),\bm{\varphi}(t),\bm{d}^{in}(t), \bm{d}^{op}(t)\},
    \vspace{-.5mm}
\end{align}
where $\gamma(t)$ represents the skewness of popularity, $\bm{h}(t)=\{h_{n}(t)\}_{n\in\mathcal{N}}$ denotes the channel gain vector for all users,  $\bm{\varphi}(t)=\{\varphi_{n}(t)\}_{n\in\mathcal{N}}$ provides information about the AIGC service requests from all users, $\bm{d}^{in}(t)=\{d^{in}_{n}(t)\}_{n\in\mathcal{N}}$ represents the input data size vector for all users, and $\bm{d}^{op}(t)=\{d^{op}_{\varphi_{n}(t)}\}_{n\in\mathcal{N}}$ is the output data size vector corresponding to the users' AIGC service requests.

\subsubsection{Action Space} After obtaining the state $\bm{s}(t)$, the corresponding action space $\bm{a}(t)$, which includes GenAI model caching, communication, and computing resource allocation decisions, is expressed as
\vspace{-1.5mm}
\begin{align}\label{eq:DDPG_action}
    \bm{a}(t)=\{ \bm{\varrho}(t),\bm{b}(t), \bm{x}(t)\},
    \vspace{-.5mm}
\end{align}
where $\bm{\varrho}(t)=\{\varrho_m(t)\}_{m \in \mathcal{M} }$ denotes the caching decision vector for all GenAI models, $\bm{b}(t)=\{b_{n}(t)\}_{n\in\mathcal{N}}$ represents the bandwidth allocation ratio vector for all users, and $\bm{x}(t)=\{x_{n}(t)\}_{n\in\mathcal{N}}$ indicates the computing resource allocation ratio vector for all users at time slot $t$.

\subsubsection{Reward Function} After executing action $\bm{a}(t)$ in response to state $\bm{s}(t)$, the environment provides feedback in the form of a reward $r(t)$, which aligns with the objective function defined in~\eqref{eq:problem}.
\vspace{-1.5mm}
\begin{align}\label{eq:DDPG_reward}
    r(t)=-\frac{1}{N}\sum_{n\in\mathcal{N}}G_{n}(t).
    \vspace{-.5mm}
\end{align}

\begin{figure}[t!]
\includegraphics[width=.48\textwidth]{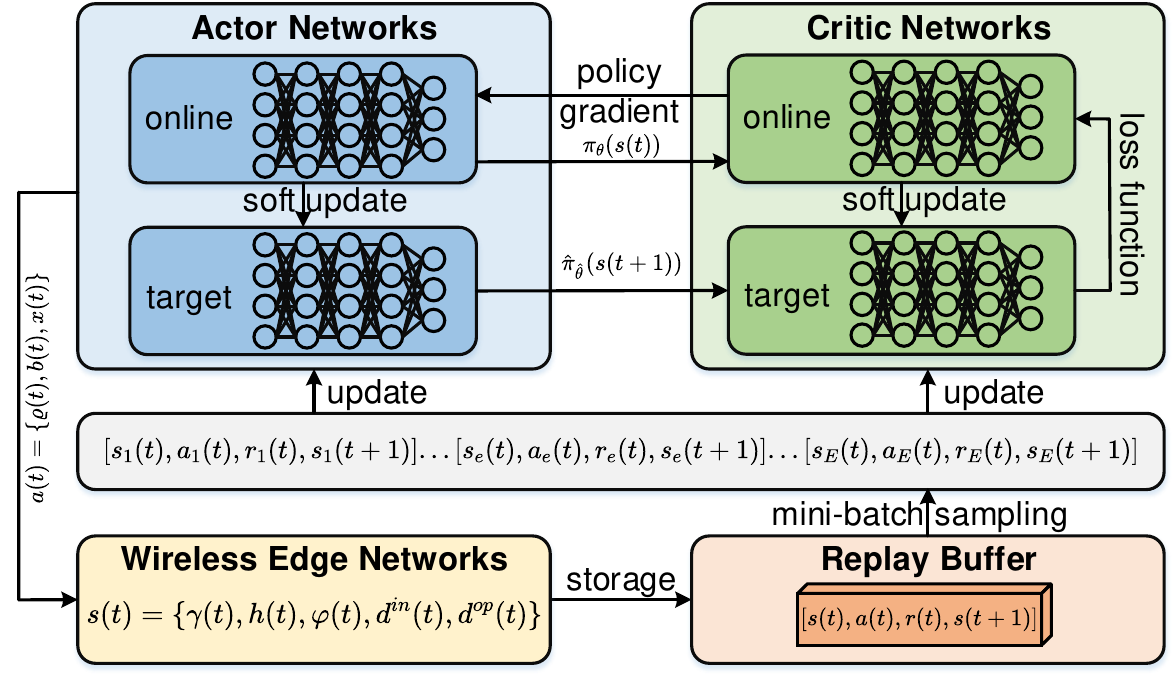}
\centering
\vspace{-1.5mm}
\caption{The architecture of the DDPG algorithm.}
\label{fig:DDPG}
\vspace{-1mm}
\end{figure}

\subsection{Architecture of the DDPG Algorithm} 
The deep deterministic policy gradient (DDPG)~\cite{lillicrap2015continuous} architecture is depicted in Fig.~\ref{fig:DDPG}, consisting of an online actor network responsible for generating actions and an online critic network for evaluating these actions. Each network is paired with respective target networks to address training instability. Additionally, a replay buffer is used to reduce sample correlation by randomly sampling stored transitions.

\subsubsection{Actor Network} The multi-layer perceptron (MLP)-based actor network $\pi_{\bm{\theta}}$, parameterized by $\bm{\theta}$, takes the state $\bm{s}(t)$ as input and outputs the action $\bm{a}(t)$. To stabilize the learning process, a target actor network $\hat{\pi}_{\bm{\hat{\theta}}}$, parameterized by $\bm{\hat{\theta}}$, is employed and structured identically to $\pi_{\bm{\theta}}$.

\subsubsection{Critic Network} The MLP-based critic network $Q_{\bm{\phi}}$, parameterized by $\bm{\phi}$, takes both state $\bm{s}(t)$ and action $\bm{a}(t)$ as inputs and outputs the Q-value function $Q_{\bm{\phi}}(\bm{s}(t),\bm{a}(t))$. Similarly, a target critic network $\hat{Q}_{\bm{\hat{\phi}}}$, parameterized by $\bm{\hat{\phi}}$, is introduced for stabilization.

\subsubsection{Replay Buffer} During training, a replay buffer $\mathcal{E}$ stores transition tuples. At each time slot $t$, DDPG stores $[\bm{s}(t), \bm{a}(t), r(t), \bm{s}(t+1)]$ in $\mathcal{E}$ for later sampling in policy development.

\subsubsection{Policy Improvement} A mini-batch of $E$ samples $\{[\bm{s}_e(t), \bm{a}_e(t), r_e(t), \bm{s}_e(t+1)]\}_{e=1}^{E}$ is randomly sampled from $\mathcal{E}$ to update the critic and actor networks. For the critic network $Q_{\bm{\phi}}$, the average temporal difference error between the target Q-value $\hat{y}_e(t)$ and the Q-value $y_e(t)$ is minimized:
\vspace{-1.5mm}
\begin{align}\label{eq:DDPG_critic}
    &\quad\quad\quad\quad  \mathbb{L}=\frac{1}{E}\sum_{e=1}^E\Big[ \frac{1}{2}(\hat{y}_e(t)-y_e(t))^2  \Big], \\
    &\hspace{-3mm}\text{s.t.}~\ y_e(t)=Q_{\bm{\phi}}(\bm{s}_e(t),\bm{a}_e(t)) \tag{15a},\\
    & \quad \hat{y}_e(t)=r_e(t)+\omega \hat{Q}_{\bm{\hat{\phi}}}(\bm{s}_{e}(t+1),\hat{\pi}_{\hat{\bm{\theta}}}(\bm{s}_{e}(t+1))), \tag{15b}
    \vspace{-.5mm}
\end{align}
where $e$ is the $e$-th transition tuple, and $\omega$ represents the discount factor. For the actor network $\pi_{\bm{\theta}}$, the sample policy gradient is updated by:
\vspace{-1.5mm}
\begin{align}\label{eq:DDPG_actor}
    \nabla_{\pi_{\bm{\theta}}}\mathcal{J}=\frac{1}{E}\sum_{e=1}^E  \nabla_{\bm{a}}Q_{\bm{\phi}}(\bm{s}_e(t),\bm{a})_{\bm{a}=\pi_{\bm{\theta}}(\bm{s}_e(t))}\nabla_{\bm{\theta}}\pi_{\bm{\theta}}(\bm{s}_e(t)).
    \vspace{-1.5mm}
\end{align}
During the training process, the parameters of the target networks are updated gradually to ensure stability in policy and Q-value changes:
\vspace{-1.5mm}
\begin{align}
    &\hat{\bm{\theta}}\leftarrow \varepsilon\bm{\theta}+(1-\varepsilon)\hat{\bm{\theta}}, \label{eq:D3PG_target_actor}\\
    &\hat{\bm{\phi}} \leftarrow \varepsilon\bm{\phi}+(1-\varepsilon)\hat{\bm{\phi}}, \label{eq:D3PG_target_critic}
    \vspace{-.5mm}
\end{align}
where $\varepsilon \in (0,1]$ is the target network update rate.
    
\section{Performance Evaluation}
\subsection{Simulation Settings}

We consider a network comprising a cloud data center, a base station (BS) functioning as an edge server, and between 10 and 18 users distributed within a 250 m × 250 m area. User mobility is captured by time-varying locations, where, at the start of each time slot $t$, each user’s location is uniformly sampled within this area. Additionally, we set the number of model types to $M=10$, with each GenAI model requiring a storage size $c_m$ within the range $[2, 10]$ GB. To simulate the varying capabilities of different GenAI models, we configure $A_1 \in [50,100]$, $A_2 \in [100,150]$, $A_3 \in [150,200]$, $A_4 \in (0,50]$, $B_1 \in (0,0.5]$, and $B_2 \in (0,10]$. Meanwhile, we model the AIGC service popularity using a finite Markov state transition model with three states: ${\gamma_1,\gamma_2,\gamma_3}$, with parameters $\gamma_1=0.2, \gamma_2 = 0.5$, and $\gamma_3=0.7$. The transition probabilities between these popularity states are set as follows:
\begin{equation} \label{eq:AIGC_service_states}
\text{Pr}^{\gamma} = 
\begin{bmatrix}
\text{Pr}_{11}^{\gamma} & \text{Pr}_{12}^{\gamma} & \text{Pr}_{13}^{\gamma} \\
\text{Pr}_{21}^{\gamma} & \text{Pr}_{22}^{\gamma} & \text{Pr}_{23}^{\gamma} \\
\text{Pr}_{31}^{\gamma} & \text{Pr}_{32}^{\gamma} & \text{Pr}_{33}^{\gamma} \\
\end{bmatrix}
=
\begin{bmatrix}
0.6 & 0.2 & 0.2 \\
0.1 & 0.7 & 0.2 \\
0.2 & 0.3 & 0.5 \\
\end{bmatrix}.
\end{equation}
The detailed settings for other parameters in our experiments are presented in Table~\ref{table:simulation_parameters}.

\begin{table}[htbp]
\vspace{-2mm}
\centering
\footnotesize
\caption{Parameters Used in Simulations ~\cite{yao2023cooperative, zhang2022two}}
\label{table:simulation_parameters}
\begin{tabular}{|l|c|}
\hline
\textbf{Parameter}                    & \textbf{Value}   \\ \hline \hline
Number of time slots $T$                  & 50                \\ \hline
Duration of each time slot $\tau$          & 20 seconds        \\ \hline
Input data size $d^{in}_{n}(t)$   & [5, 10] MB           \\ \hline
Uplink bandwidth $W^{up}$                  & 20 MHz                \\ \hline
Downlink bandwidth $W^{dw}$                  & 40 MHz                \\ \hline
Transmit power of user $p_n$                  & 23 dBm        \\ \hline
Transmit power of the BS $p_b$                  & 43 dBm        \\ \hline
Noise power spectral density $U_0$                  & -176 dBm/Hz        \\ \hline
Backhual transmission rate $R^{bc}=R^{cb}$                  & 100 Mbps          \\ \hline
Output data size $d^{op}_{m}$              & [5, 10] MB           \\ \hline
Number of denoising steps performed at the BS $L$         & 1000                \\ \hline
Storage capacity of the BS $C$    & 20 GB             \\ \hline
Preference weight factor $\alpha$  & 0.7              \\ \hline
Reward discount factor $\omega$                  & 0.99          \\ \hline
Target network update rate $\varepsilon$                          & 0.005        \\ \hline
Number of episodes $H$                          & 500        \\\hline
\end{tabular}
\end{table}

\vspace{-3mm}
\subsection{Benchmark Solutions}
To demonstrate the effectiveness of the DDPG algorithm, we compare it with two benchmark solutions:

\begin{itemize}
    \item \emph{Heuristic-based caching and resource allocation scheme (HCRAS):} At each time slot, model caching, bandwidth, and computing resource allocation decisions are generated using a genetic algorithm. Specifically, HCRAS employs real-valued encoding to create multiple chromosomes (solutions) that form the initial population. These chromosomes evolve through simulated binary crossovers and polynomial mutations until the maximum iteration limit is reached. Finally, the chromosome with the lowest value in~\eqref{eq:problem} is selected.
    \item \emph{Randomized caching and average resource allocation scheme (RCARS):} At each time slot, the BS randomly caches GenAI models until the caching capacity is met, while bandwidth and computing resources are equally distributed among users. This approach establishes a lower-bound baseline for evaluating performance.
\end{itemize}

\begin{figure*}[htbp]
	\centering
	\begin{minipage}{0.325\linewidth}
		\centering
		\includegraphics[width=\linewidth]{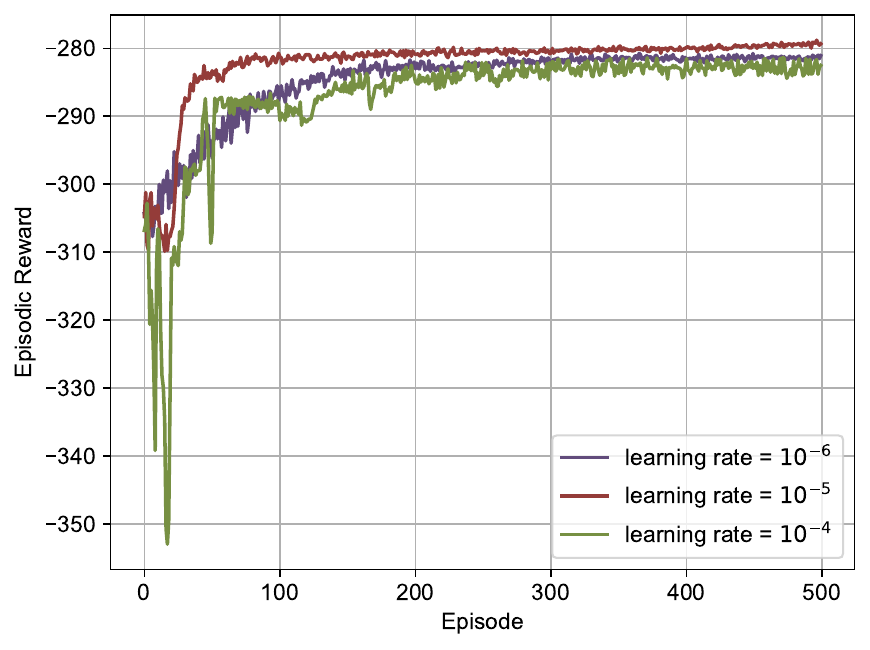}
		\caption{Learning rate impact.}
		\label{fig:DDPG_convergence_different_LR}
	\end{minipage}
	\begin{minipage}{0.333\linewidth}
		\centering
		\includegraphics[width=\linewidth]{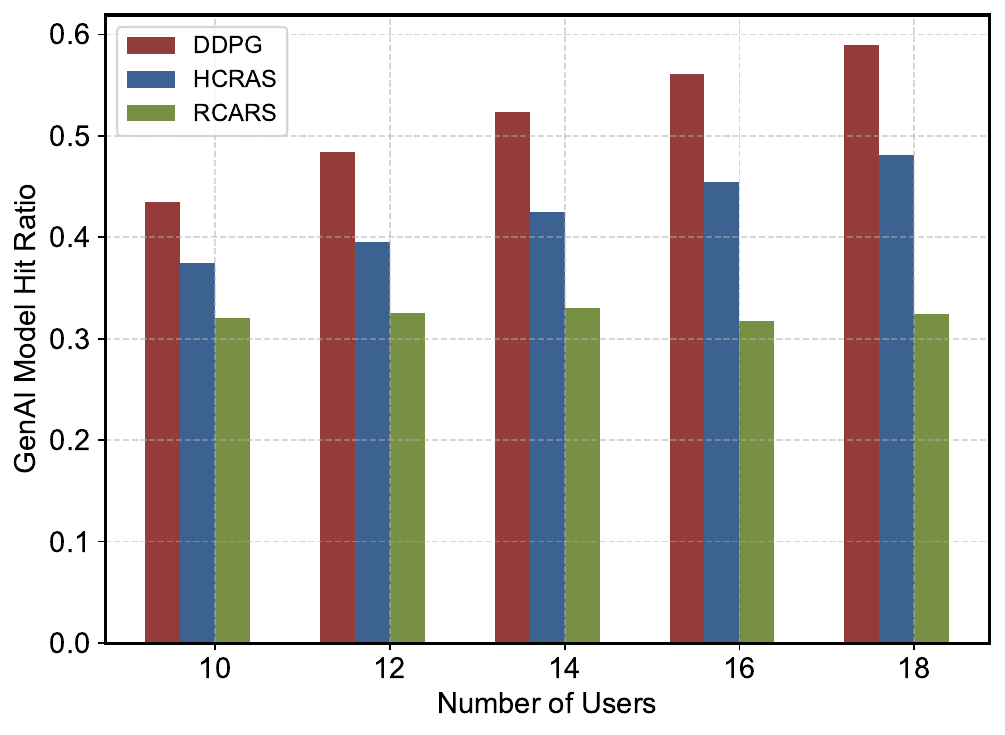}
		\caption{User number impact.}
		\label{fig:Model_hit_ratio_different_users}
	\end{minipage}
    \begin{minipage}{0.318\linewidth}
		\centering
		\includegraphics[width=\linewidth]{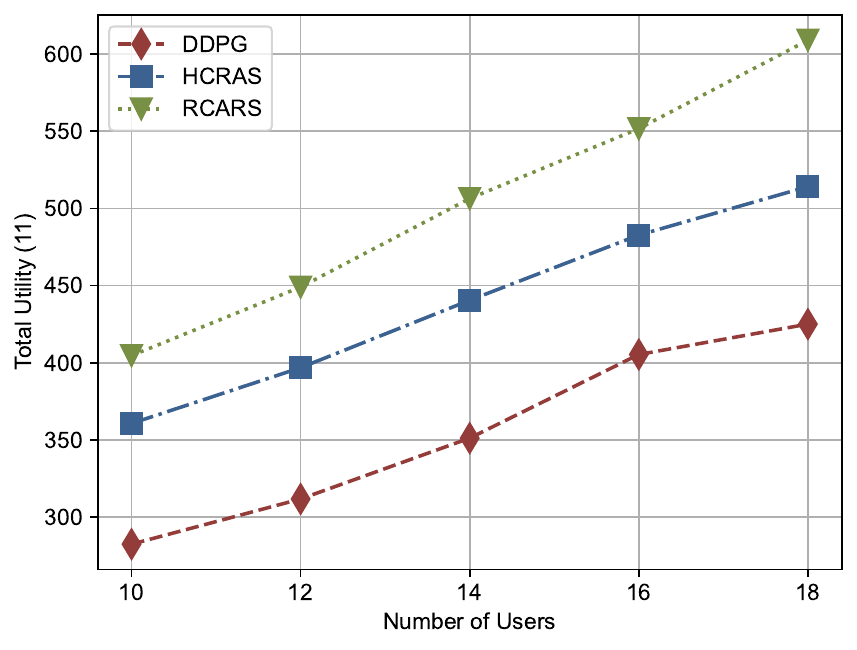}
		\caption{User number impact.}
		\label{fig:Objective_value_different_users}
	\end{minipage}
\vspace{-6mm}
\end{figure*}

    \subsection{Simulation Results}
    We illustrate the convergence behavior of the DDPG algorithm under different learning rates. As shown in Fig.~\ref{fig:DDPG_convergence_different_LR}, the converged reward initially rises but then declines as the learning rate continuously increases. This pattern occurs because a moderate increase in the learning rate helps prevent the DDPG agent from becoming trapped in local optima. However, further amplification of the learning rate may exacerbate temporal difference errors, leading to biased value estimations. 

    The results in Fig.~\ref{fig:Model_hit_ratio_different_users} depict the impact of increasing the number of users from 10 to 18 on the GenAI model hit ratio. As the user count grows, the model hit ratio improves. This is because the distribution of AIGC service requests increasingly aligns with the underlying probability distribution, with popular AIGC services comprising a larger share of the total requests. Overall, DDPG demonstrates superior performance compared to the other algorithms. Its
    performance is 15.99\% better than HCRAS, and 35.93\% better than RCARS at 10 users; 22.58\% better than HCRAS, and 81.87\% better than RCARS at 18 users.

    Fig.~\ref{fig:Objective_value_different_users} also shows how the number of users affects the total utility (\ref{eq:problem}). As the user count rises, the total utility (\ref{eq:problem}) increases as well. This is due to heightened competition among users for a fixed amount of bandwidth and computational resources, resulting in fewer resources per user. In this regard, DDPG continues to outperform the other algorithms. Its performance is 21.69\% better than HCRAS, and 30.19\% better than RCARS at 10 users; 17.33\% better than HCRAS, and 30.21\% better than RCARS at 18 users.

    \section{Conclusion}
    In this paper, we tackled the challenge of joint model caching and resource allocation in generative AI-enabled wireless edge networks, with the goal of delivering low-latency, high-quality AIGC services. To this end, we first formulated the problem as an MDP and employed a DDPG algorithm to optimize discrete (i.e., GenAI model caching decisions) and continuous variables (i.e., bandwidth allocation and denoising step allocation decisions) concurrently. Experimental results demonstrate that the DDPG algorithm not only achieves a higher model hit ratio but also provides superior quality and lower latency for AIGC services.

    \bibliographystyle{IEEEtran}
    \bibliography{main.bbl}

    \end{document}